\documentclass[a4paper]{article}
\usepackage[ansinew]{inputenc}
\usepackage{times}
\usepackage[T1]{fontenc}
\usepackage{graphicx}
\usepackage{geometry}
\usepackage{amssymb}
\usepackage{amsmath}

\usepackage{rotating}

\usepackage{xcolor}
\colorlet{shadecolor}{gray!25}
\usepackage{subfig}
\usepackage{float}
\usepackage{lscape}

\author{Ludwig A. Hothorn,\\ 
Im Grund 12, D-31867 Lauenau, Germany\\ \scriptsize(retired from Leibniz University Hannover)}

\title{Comparisons of proportions in $k$ dose groups against a negative control assuming order restriction: Williams-type test vs. closed test procedures}
\begin{document}

\maketitle
\begin{abstract}
The comparison of proportions is considered in the asymptotic generalized linear model with the odds ratio as effect size. 
When several doses are compared with a control assuming an order restriction, a Williams-type trend test can be used. As an alternative, two variants of the closed testing approach are considered, one using global Williams-tests in the partition hypotheses, one with pairwise contrasts. Their advantages in terms of power and simplicity are demonstrated. Related R-code is provided.
\end{abstract}

\section{The problem}\label{sec1}
When you think of 'proportions', 'k-sample design' and 'order restriction', the first thing that comes to mind is the Armitage trend test \cite{ARMITAGE1955}, which is one of the most cited statistical tests. As the title of Armitage's poineering work suggests, it is constructed for a linear trend in proportions, but one wants to be sensitive to arbitrary shapes of monotonic alternatives. Furthermore, it does not specifically compare to a negative control, is only a global test (i.e. does not provide elementary
information to $D_i-C$) and models the dose levels as a quantitative covariate (in the sense of a linear logistic regression).
The simultaneous comparison of proportions with a control or placebo is not so often described in the biomedical literature (compared to continuous endpoints). For example, the  comparison of the overall response rates of three doses of oral liarozole  with placebo in a randomized dose-finding trial for the treatment of psoriasis was performed by the approach \cite{PIEGORSCH1991} without using order restriction \cite{BerthJones2000}. Comparisons of multiple dose groups with a control group assuming a monotonic dose-response relationship are frequently performed by means of the Williams trend test \cite{Williams1971} where a modification for risk differences (of proportions) is available \cite{Hothorn2010}. 

Motivation to use the Williams test instead of the Dunnett test \cite{Dunnett1955} (without assuming order restriction) is increased power (by restricting $H_1$) and the superior interpretation of a trend (both global and selected parts of the dose-response relationship). The main difference between Dunnett and William's test is that the former considers comparisons between $C$ and the individual $D_i$, but the latter does not consider  the comparison with explicit doses but pooled doses (except for $D_{max}-C$). Therefore, an order restricted test is derived here for comparison to control with the individual doses.\\
The closed testing procedure (CTP) \cite{MARCUS1976} is an alternative to the max-T test for multiple contrasts on which the William test is based \cite{Bretz2006}. Two special cases are used in the following: the complete family of hypotheses when comparing to control only \cite{Sonnemann2008} and the decision tree reduction when assuming order restriction \cite{Hothorn1997}. Therefore, related closed testing versions of order restricted tests are derived here. \\

For proportions three effect sizes, risk difference, risk ratio and odds ratio are available. Because approximate glm-type approaches are used here, the odds ratio is considered as an example.

\section{The Williams procedure}
The Williams test can be formulated as multiple contrast test (MCT) \cite{SHAFFER1977}, \cite{Mukerjee1987}. A maxT- test is used for proportions $\pi_i$ and their ML-estimates $p_i$: $t_{MCT}=max(t_1,...,t_{q'})$ with $	t_q=\sum_{i=0}^k c_i\bar{p}_i/S \sqrt{\sum_i^k c_i^2/n_i}$ where $c_i^q$ are the contrast coefficients (see below). The common-used adjusted p-values are given by the minimum empirical $\alpha$-level: $\frac{\sum_{i=0}^k c_i\bar{p}_i}{S\sqrt{\sum_i^k c_i^2/n_i}} = t_{q,df,R,1-sided,1-min(\alpha)}$ where $t_{q,df,R,1-sided,1-\alpha}$ is the quantile of central q-variate t distribution, available in the package mvtnorm \cite{Mi2009}. Compatible to the adjusted p-values are (two) or one-sided lower simultaneous confidence limits which are not considered here because of their difficulties in the CTP \cite{Guilbaud2018}.\\
Of course, it is also possible to model the dose as a quantitative covariate, using the Armitage trend test for near-linear profiles or the Tukey trend test for any shape \cite{Schaarschmidt2020}.
\section{Closed testing procedures}
First, the interesting elementary hypotheses $H_i: \pi_i-\pi_0$ are defined, followed by a decision tree containing all subset intersection hypotheses up to the global hypothesis, involving these elementary hypotheses \cite{MARCUS1976}. One rejects $H_i$ at level $\alpha$ if and only if $H_i$ itself is rejected and all hypotheses which include them (each at level  $\alpha$). Each hypothesis is tested with a level $\alpha$-test, with any appropriate test - this allows a high flexibility of the here described approach. Each of these tests (determined by the $\xi$ elementary hypotheses) is an intersection-union test (IUT), i.e. $T^{CTP}=min(T_1,...,T_{\xi})$, or more common $p^{CTP}=max(p_1,...,p_{\xi})$. In general CTP, the subset hypotheses can be complex and contradictory, but when considering hypotheses for comparisons with a control, they form a simple, so-called complete family of hypotheses \cite{Sonnemann2008}.
For the simple design with $k=2$, the family include the elementary (e.g. $H_0^{01}$), intersection (e.g. $H_0^{012}$) and global hypotheses (e.g. $H_0^{0123}$): \\
$H_0^{01}: \pi_0=\pi_1 \subset [H_0^{012}, H_0^{013}] \subset H_0^{0123}$\\
$H_0^{02}: \pi_0=\pi_2 \subset [H_0^{012}, H_0^{023}] \subset H_0^{0123}$\\
$H_0^{03}: \pi_0=\pi_3 \subset [H_0^{013}, H_0^{023}] \subset H_0^{0123}$\\

Monotonic order restriction $H_1: \pi_0\leq \pi_1\leq...\leq\pi_k|\pi_0<\pi_k$ (for any possible pattern of equalities/inequalities) further simplified the CTP vs. control seriously. Under this restriction the rejection of $H_0^{0123}$ implies the rejection of $H_0^{013}$ and $H_0^{03}$ and the rejection of $H_0^{012}$ implies the rejection of $H_0^{02}$. Thus the hypothesis system is significantly simplified (in the above $k=2+1$ example):\\
$H_0^{01}: H_0^{01} \wedge H_0^{012} \wedge H_0^{0123}$\\
$H_0^{02}: H_0^{012} \wedge H_0^{0123} $\\
$H_0^{03}: H_0^{0123}$\\

Any level $\alpha$ test can be used for these hypotheses. The elementary hypotheses are tested by contrast tests for $\pi_i-\pi_0$, not by 2-sample tests. For the partition and global hypotheses any order-restricted test can be used. Because of comparing $D_i-C$ two versions are considered here: i) Williams global test for each subset (denoted as C), ii) contrast tests for $\pi_{\xi}-\pi_0$ (where $\xi$ is the highest dose in the particular subset) (denoted as P).\\
For a similar objective the complete closure test for all pair comparison elementary hypotheses without order restriction and subset omnibus heterogeneity tests was considered \cite{Lehmann2018a}.

\section{Simulation study}

The power of  the tests are compared by a simulation study for a low-dimensional balanced one-way layout.
Random experiments with a single primary proportion $p_{i}, k=2$ were used to estimate the per-pairs power $\Pi_{01}, \Pi_{02},\Pi_{03}$ for several strict monotonic alternatives and two shapes with a downturn effect at the high dose  as well as the empirical FWER estimate under global and partial $H_0$ based on 5000 samples drawn independently from binomial distributions $Bin(n_i; \pi_i)$. Common simulation studies on MCTs compare the any-pair power \cite{Hothorn2020} or average power \cite{Stevens2017}. These concepts simplifies power comparisons considerably, but is not target-oriented, since which particular comparison is in the alternative is not considered. But one does not want to know if any dose from the negative control. No, you want to evaluate exactly a particular dose  relative to control (see the motivating examples above). Therefore the concept of per-pairs power is used here, although it is difficult to interpret (and therefore k=2 was used).\\
The four tests are abbreviated with $D_i$ (Dunnett original), $W_3$ (Williams-type, for the comparable $W_3$ only), $C_i$ (CTP using subset Williams-type global tests) and $P_i$ (CTP using pairwise contrasts) (with $D,W,C,P^a$ the any-pairs power). Instead complete power curves, only two relevant points in the alternative are considered for $\Pi_3>0.8$ and $\Pi_3>0.9$ .

\begin{table}[ht]
\centering\tiny
\begin{tabular}{l|l|rrrr|rrrr|rr|rrrr|rrrr}
  \hline
H	 & $n_i$ & $\pi_1$ & $\pi_2$ & $\pi_3$ & $\pi_4$ & $D_1$ & $D_2$ & $D_3$ & $D^a$ & $W_3$ & $W^a$ & $P_1$ & $P_2$ & $P_3$ & $P^a$ & $C_1$ & $C_2$ & $C_3$ & $C^a$ \\ 
  \hline

$H_0$	 & 50 & .05 & .05 & .05 & .05 & .002 & .001 & .001 & .004  & .004 & .000 & .001 & .010 & .010 & .000 & .000 & .004 & .004 \\ 
 & 50 & .07 & .07 & .07 & .07 & .005 & .003 & .004 & .011 & .002 & .012 & .001 & .002 & .018 & .018 & .001 & .003 & .012 & .012 \\ 
 & 50 & .10 & .10 & .10 & .10 & .008 & .008 & .009 & .022 & .008 & .024 & .002 & .004 & .028 & .028 & .005 & .010 & .024 & .024 \\ 
 & 50 & .20 & .20 & .20 & .20 & .013 & .018 & .016 & .038 & .018 & .037 & .005 & .011 & .041 & .041 & .011 & .019 & .036 & .036 \\ \hline
 & u & .10 & .10 & .10 & .10 & .019 & .028 & .046 & .083 & .022  & .081 & .004 & .015 & .104 & .104 & .016 & .033 & .081 & .081 \\ 

\hline\hline
$H_1$ & 50 & .05 & .05 & .05 & .30 & .002 & .002 & .877 & .877 & .910 &  & .000 & .010 & .949 & .949 & .000 & .007 & .910 & .910 \\ 
      & 50 & .05 & .10 & .20 & .30 & .056 & .496 & .895 & .921 & .934 &  & .102 & .687 & .961 & .961 & .109 & .634 & .957 & .957 \\ 
      & 50 & .05 & .30 & .30 & .30 & .904 & .903 & .914 & .993 & .931 &  & .886 & .916 & .958 & .958 & .951 & .983 & .995 & .995 \\ 
      & 50 & .05 & .05 & .10 & .30 & .002 & .048 & .885 & .886 & .922 &  & .005 & .118 & .958 & .958 & .006 & .091 & .921 & .921 \\ 
			& 50 & .05 & .10 & .30 & .20 & .054 & .892 & .504 & .915 & .608 &  & .112 & .673 & .687 & .687 & .126 & .869 & .887 & .887 \\ 
      & 50 & .05 & .10 & .30 & .10 & .055 & .892 & .054 & .893 & .091 &  & .036 & .135 & .136 & .136 & .110 & .572 & .573 & .573 \\ 
   \hline   \hline
$H_1$	& 50 & .07 & .07 & .07 & .30 & .002 & .005 & .790 & .791 & .840 &  & .002 & .022 & .894 & .894 & .002 & .011 & .840 & .840 \\ 
      & 50 & .07 & .07 & .10 & .30 & .006 & .022 & .798 & .798 & .855 &  & .006 & .074 & .906 & .906 & .010 & .052 & .855 & .855 \\ 
      & 50 & .07 & .30 & .30 & .30 & .821 & .836 & .822 & .968 & .854 &  & .787 & .840 & .908 & .908 & .882 & .944 & .969 & .969 \\ 
      & 50 & .07 & .10 & .30 & .20 & .032 & .791 & .347 & .820 & .448 &  & .069 & .533 & .552 & .552 & .081 & .751 & .779 & .779 \\ 
   \hline
   \hline
\end{tabular}
\caption{Per-power estimates $\Pi_i$ for selected alternatives}
\end{table}

Per definition all tests control the FWER in a weak sense (i.e. all elementary hypotheses are under $H_0$ and in a strong sense, only a particular elementary hypothesis is under $H_0$ considering order restriction. As expected, these asymptotic tests do not control FWER when $\pi$ and $n_i$ are rather lower, particularly in unbalanced design (u). For strict monotonic alternatives, the power of the Williams test is per definition slightly larger than that of the Dunnett test (only directly comparable for $D_3-0$). Both closed tests almost always show a power superiority for all $\Pi_i$, for some patterns a clear superiority, compared to the Dunnett test.  For non-monotonic shapes, depending on the amount of decline at $D_{max}$ (downturn effect), all tests assuming an order restriction are not robust as expected.

\section{Evaluation of the real data example}
The overall percentage response rates in the liarozole trial  are $6, 18,11,38 \%$ in the randomized dose groups $D_i=0,50,75,150$ mg/kg with $n_i=34,35,36,34$ subjects \cite{BerthJones2000}. Table 2 shows the adjusted p-values for increasing the odds ratios versus control in the three dose group. In this data example the closed testing procedure using conditional pairwise contrasts (P) reveals the smallest p-value for the $150-0$ comparison. The related R-code is given in the Appendix.

\begin{table}[ht]
\centering\small
\begin{tabular}{r|l|l|l|l}
  \hline
 $H_{0,i}$ & Dunnett & Williams & P & C  \\ 
  \hline
 50 - 0 & 0.153 			& ... 		& 0.221 	& 0.153  \\ 
 75 - 0 & 0.362 		& ... 		& 0.221 	& 0.153  \\ 
 150 - 0 & 0.0056 	& 0.0036 	& 0.0023 	& 0.0036 \\ 
   \hline
\end{tabular}
\caption{One-sided adjusted p-values for the 4 test procedures in the liarozole trial (P... CTP using pairwise contrasts, C... CTP using global Williams trend tests in the subsets)} 
\end{table}

\section{Conclusions}
No ump-test exists for any pattern of monotonic $H_1$, certainly not for alternatives with downturns at high dose(s). Considering the particularly pattern as unknown a-priori, the here proposed CTP's can be recommended for the analysis of proportions in a k-sample design assuming order restriction, where the odds ratio is used as effect size in a generalized linear model.

\section{Appendix}
\scriptsize
\begin{verbatim}
Dose <-c(rep("0", 34), rep("1", 35),rep("2", 36),rep("3", 34))
lia <-c(rep("no",32), rep("resp",2), rep("no",29), rep("resp",6),
          rep("no",32),rep("resp",4),rep("no",21),rep("resp",13))
Li <-data.frame(lia,Dose)
modLi <- glm(lia ~ Dose, data=Li, family= binomial(link="logit"))
library("multcomp")
Du=summary(glht(modLi, linfct = mcp(Dose = "Dunnett"),alternative="greater"))$test$pvalues
Wi=summary(glht(modLi, linfct = mcp(Dose = "Williams"),alternative="greater"))$test$pvalues
Wi0123=min(summary(glht(modLi, linfct = mcp(Dose = "Williams"),alternative="greater"))$test$pvalues)
contMat03<-c(-1,0,0,1); contMat02<-c(-1,0,1,0); contMat01<-c(-1,1, 0,0)
n012<-c(34,35,36)
cmWi012<-contrMat(n012, type="Williams");  contr4 <-c(0,0);   cWi012<-cbind(cmWi012,contr4)
Wi012=min(summary(glht(modLi, linfct = mcp(Dose=cWi012),alternative="greater"))$test$pvalues)
C03=summary(glht(modLi, linfct = mcp(Dose = contMat03),alternative="greater"))$test$pvalues
C02=summary(glht(modLi, linfct = mcp(Dose = contMat02),alternative="greater"))$test$pvalues
C01=summary(glht(modLi, linfct = mcp(Dose = contMat01),alternative="greater"))$test$pvalues

pa01=max(C01,C02,C03)
pa02=max(C02,C03)
pa03=max(C03)# CTP pairwise
WW01=max(C01,Wi012,Wi0123) # CTP Williams all subset
WW02=max(Wi012,Wi0123)
WW03=Wi0123

\end{verbatim}

\footnotesize

\bibliographystyle{plain}
\footnotesize

\end{document}